\documentclass[24pt]{article}
\usepackage{amssymb}
\usepackage{epsfig,caption,graphicx,eepic,epic}
\usepackage{pstricks}
\usepackage{amssymb,amsmath}
\usepackage{multido}
\usepackage{array}
\begin{document}
\def\be{\begin{equation}}
\def\ee{\end{equation}}
\def\ba{\begin{eqnarray}} 
\def\ea{\end{eqnarray}}
\def\nn{\nonumber}
\newcommand{\bbf}{\mathbf}   
\newcommand{\rrm}{\mathrm}
\title{\bf Coherent and decoherent time evolution of finite Markovian and non-Markovian open quantum systems}
\author{Tarek Khalil$^{a,}$
\footnote{E-mail address: tkhalil@ul.edu.lb}\\ 
and\\
Jean Richert$^{b}$
\footnote{E-mail address: j.mc.richert@gmail.com}\\ 
$^{a}$ Department of Physics, Faculty of Sciences(V),\\
Lebanese University, Nabatieh,
Lebanon\\ 
$^{b}$ Institut de Physique, Universit\'e de Strasbourg,\\
3, rue de l'Universit\'e, 67084 Strasbourg Cedex,\\      
France} 
 
\date{\today}
\maketitle 

\begin{abstract}
We examine the properties of open quantum systems with respect to their time evolution in different regimes, Markovian and non-Markovian. We analyze their behaviour with respect to their coherent or decoherent time evolution by means of different models and try to gain some insight into the possible correlations between Markovianity and coherence. 
\end{abstract} 
\maketitle  



\section{Introduction}

The properties of open quantum systems has a long history which starts with the measure problem already at the eve of the existence of quantum mechanics \cite{neu,zeh,whe,zur1,per,kie,zur2,sch,lan,bal}. The question of the measure of a quantum observable with a device which works on a classical level is related to the correlated concepts of decoherence and irreversibility. Although much progress has been made there remain many open questions leading to critics and controversies on the subject~\cite{kas}. An answer to these questions is of a particularly strong interest nowadays to the community which develops an intense activity in the framework of new fields like information theory and quantum computing.

The central property of an open quantum system is the characteristic behaviour of its time evolution. There are two aspects concerning this point. The first aspect is linked to the length of the time interval over which the system keeps the memory of its interaction with its environment. If this interval is short compared to the proper characteristic evolution time of the system the memory of the interaction is quickly lost, the regime is called Markovian. If this is not the case memory effects induced by a finite time response of the environment are present, so called backflow effects are generated, the process is called non-Markovian.

The second aspect which is closely related to the measure problem concerns the evolution of the components of the density operator in a fixed basis of states. Given a specific basis of states of the system the evolution can either lead to so called decoherence if the non diagonal matrix elements of this operator decrease to zero, coherent if this is not the case. The question is whether and how the Markovian or non-Markovian property is correlated to a coherent or decoherent behaviour of the open system. 

Time behaviour and structure of the system-environment complex are necessarily intimately linked~\cite{kr}. In fact
Markovianity (non-Markovianity) and coherence (decoherence) are necessarily generated by the physical properties of the system, i.e. its environment and the interaction which couples them dynamically. This central point appears in the background of many different studies but has never been, to our knowledge, systematically analyzed as such. In the present work we aim to consider this point by means of models which allow to show how the structure of the different components of the total system determine the time evolution  of the open system.  

The structure of the paper is the following. In section 2 we introduce a general formal expression of the density operator. In section 3 we develop two Markovian systems of different types which show different time evolutions.
Section 4 is devoted to a non-Markovian system and in section 5 we show explicitly how the algebraic structures of the Hamiltonian of the environment and the coupling interaction between the system and the environment determine the behaviour  of the non-diagonal elements of the density operator. Section 6 shows explicitly how the spectral 
properties of the environment can fix the transition from a Markovian to a non-Markovian behaviour and its consequence with regard to the type of evolution, either coherent or decoherent. Section 7 summarizes and analyzes the results. Details of calculations are developed in Appendices.

\section{The density operator}

We consider an open quantum system $S$ characterized by a density operator  $\hat\rho_{S}(t)$ which evolves in time 
from $t_{0}$ to $t$ under the action of the evolution operator $\hat T(t,t_{0})$ 

\ba
\hat\rho_{S}(t)=\hat T(t,t_{0})\hat\rho_{S}(t_{0})
\label{eq1}
\ea

At the initial time $t_{0}$ the system $S$ is supposed to be decoupled from its environment $E$ and characterized by the density operator

\ba
\hat\rho_{S}(t_{0})= \sum_{i_{1},i_{2}}c_{i_{1}}c_{i_{2}}^{*}|i_{1}\rangle \langle i_{2}|
\label{eq2}
\ea
and in $E$ space the density operator $\hat\rho_{E}$ reads
\ba
\hat\rho_{E}(t_{0})= \sum_{\alpha_{1},\alpha_{2}}d_{\alpha_{1},\alpha_{2}}|\alpha_{1}\rangle \langle \alpha_{2}|
\label{eq3}
\ea

Here $|i_{1}\rangle, |i_{2}\rangle $ and $|\alpha_{1}\rangle, |\alpha_{2}\rangle$ are orthogonal states in  
$S$ and $E$ space respectively, $c_{i_{1}},c_{i_{2}}$ normalized amplitudes and $d_{\alpha_{1},\alpha_{2}}$ weights such that $\hat\rho_{E}^{2}(t_{0})=\hat\rho_{E}(t_{0})$.

At time $t>t_{0}$ the reduced density operator in $S$ space is $\hat\rho_{S}(t)=Tr_{E}[\hat\rho(t)]$ where $\hat\rho(t)$ is the density operator of the total system $S+E$. It can be written as~\cite{vlb}

\ba
\hat\rho_{S}(t)=\sum_{i_{1},i_{2}}c_{i_{1}}c_{i_{2}}^{*}\hat\Phi_{i_{1},i_{2}}(t,t_{0})
\label{eq4}
\ea

with

\ba
\hat\Phi_{i_{1},i_{2}}(t,t_{0})=\sum_{j_{1},j_{2}}C_{(i_{1},i_{2}),(j_{1},j_{2})}(t,t_{0})|j_{1}\rangle \langle j_{2}|
\label{eq5}
\ea

where the super matrix $C$ reads

\ba
C_{(i_{1},i_{2}),(j_{1},j_{2})}(t,t_{0})=\sum_{\alpha_{1},\alpha_{2},\gamma}d_{\alpha_{1},\alpha_{2}}
U_{(i_{1}j_{1}),(\alpha_{1}\gamma)}(t,t_{0}) U_{(i_{2}j_{2}),(\alpha_{2}\gamma)}^{*}(t,t_{0})
\label{eq6}
\ea
with 

\ba
U_{(i_{1}j_{1}),(\alpha_{1}\gamma)}(t,t_{0})=\langle j_{1}\gamma|U(t,t_{0})|i_{1} \alpha_{1}\rangle
\notag\\
U^{*}_{(i_{2}j_{2}),(\alpha_{2}\gamma)}(t,t_{0})=\langle i_{2} \alpha_{2}|U^{+}(t,t_{0})|j_{2} \gamma \rangle
\label{eq7}
\ea
 
The unitary evolution operator reads $U(t,t_{0})=e^{-i\hat H(t-t_{0})}$ where $\hat H$ is the total Hamiltonian in 
$S+E$ space and the super matrix $C$ obeys the condition 
$\lim_{t\rightarrow t_{0}}C_{(i_{1},i_{2}),(j_{1},j_{2})}(t,t_{0})=\delta_{i_{1},i_{2}} \delta_{j_{1},j_{2}}$.\\

In the present formulation the system is described in terms of pure states. The results which will be derived below 
remain valid if the initial density operator at the initial time is composed of mixed states 
$\hat\rho_{S}(t_{0})=\sum_{i_{1},i_{2}}c_{i_{1}i_{2}}|i_{1}\rangle \langle i_{2}|$.\\

We shall now use the general expression given by Eq.(4) in order to work out the density operator and matrix elements of $\hat\rho_{S}(t)$ in the framework of different systems in order to characterize their time evolution properties with respect to Markovianity and coherence.\\ 

\section{Two Markovian systems}

The most general Hamiltonian acting in $S+E$ space can be decomposed as $\hat H=\hat H_{S}+\hat H_{E}+\hat H_{SE}$
where $\hat H_{SE}$ generates the interaction between $S$ and $E$.

It has been shown that the Markovian or non-Markovian property of $S$ can be characterized by the presence of a unique state in $E$ space~\cite{sti,riv} or, more generally, by the commutator $\hat C_{H} \equiv [\hat H_{E},\hat H_{SE}]$ ~\cite{kr}. The evolution of $S$ will be Markovian if $\hat C_{H}=0$ and non-Markovian if $\hat C_{H} \not= 0$.
The consequences of these structures are analyzed by means of the following models.

\subsection{Model 1: The environment space reduces to a single state}

The Hamiltonian of the total system $\hat H= \hat H_{S}+\hat H_{E}+\hat H_{SE}$ in this order reads

\ba
\hat H=\omega_{S} \hat J_{z}+\beta a^{+}a+\eta(a^{+}\hat J_{-}+a\hat J_{+})
\label{eq8}
\ea
where $\hat J$ and $\hat J_{z}$ are the angular momentum operator and its projection on the $Oz$ and 
$\hat J_{-}=\hat J_{x}-i\hat J_{y},\hat J_{+}= \hat J_{x}+i\hat J_{y}$. The operators $a^{+}$ and $a$ create and annihilate bosonic quanta, $\beta$ and $\eta$ are real parameters. 

Using the Zassenhaus expansion ~\cite{zas,ca} of the propagators $\hat U(t,t_{0})$ and $\hat U^{+}(t,t_{0})$ which appear in the expression of the density operator it is possible to develop the exponential in terms of the Hamiltonian operators, see Appendix A. The expression of $\hat H$ is developed into its components and use in made of the commutation relations obeyed by the components of the angular momentum operator $\hat J$. The development consists of an infinite series of terms which cannot be worked out in practice. 

However the structure of the development can be obtained. This is possible because there is only one state 
$|\alpha \rangle$ in $E$ space. In this case the only contributions to the matrix elements of $\hat U, \hat U^{+}$ 
are the exponential terms which contain $a^{+}a, \hat J_{z}$ and $ \hat J_{-}\hat J_{+}$ as well as their combinations.\\

As a consequence the matrix elements $\hat \rho_{S}(t,t_{0})$ are also diagonal in $S$ space since the angular 
momentum operators defined above lead to diagonal matrix elements in the natural basis of states $|jm\rangle$
which are eigenvectors of $\hat J^{2}$ and $\hat J_{z}$. The expression reads

\ba
\rho_{S}^{(jj)}(t,t_{0})=|c_{j}|^{2}|u_{j \alpha}(t,t_{0})|^{2}
\label{eq9}
\ea
where $u_{j \alpha}(t,t_{0})=\langle j \alpha|U(t,t_{0})|j \alpha \rangle$. In this example it comes out that there 
are no non-diagonal elements in the expression of the density operator whatever the size of the system space $S$,
hence no decoherence because of the structure of the environment. In the sequel $t_{0}=0$.

\subsection{Model 2: A Markovian system}    

We consider another Markovian system for which $[\hat H_{E},\hat H_{SE}] = 0$ in order to test its property with respect to coherence. Here

\begin{center} 
\ba
\hat H_{S}= \omega_{S} a^{+}a
\notag \\
\hat H_{E}=\omega \hat J_{z}
\notag \\
\hat H_{SE}=\eta(a^{+}+a) \hat J^{2}
\label{eq10}
\ea 
\end{center}  
 
Using the corresponding notations introduced above the density operator $\hat \rho_{S}(t)$ reads 

\ba
\hat \rho_{S}(t)=\sum_{n_{1},n_{2}}\sum_{n_{3},n_{4}}C_{(n_{1}n_{2})(n_{3}n_{4})}(t)c_{n_{1}}c^{*}_{n_{2}}
|n_{3} \rangle \langle n_{4}|
\label{eq11}
\ea 
where the $|n_{i}\rangle$'s are now the normalized eigenstates of $\hat H_{S}$. The super matrix $C$ contains now the trace over the states [$|j_{i} m_{i}\rangle$] in $E$ space which are eigenstates of $\hat J_{z}$ and $\hat J^{2}$,
$\hat J^{2}|j_{i} m_{i}\rangle=j_{i}(j_{i}+1)|j_{i} m_{i}\rangle$. 

Since 
$\hat H_{S}$ is diagonal in this basis of states $C$ now reads

\ba
C_{(n_{1}n_{2})(n_{3}n_{4})}(t)=\sum_{j m}\langle n_{3 } jm|U(t)|n_{1} jm \rangle  
\langle n_{2} jm |U^{*}(t) |n_{4} jm\rangle/\hat j^{4}
\label{eq12}
\ea  
and $C_{(n_{1}n_{2})(n_{3}n_{4})}(0)=1$.

Using the decomposition of the evolution operator $e^{-it\hat H}$ already used in model 1 the projection on $E$ space leads to
  
\ba
\langle jm|e^{-it\hat H}|jm\rangle=e^{-it\omega_{s} a^{+}a}e^{-i\eta j(j+1)\sin(\omega_{s}t)(a^{+}+a)/\omega_{s}}
e^{\omega_{s}[1-\cos(\eta j(j+1)t](a^{+}-a)}
\label{eq13}
\ea   
 
The expression of $C_{(n_{1}n_{2})(n_{3}n_{4})}(t)$ can now be explicitly worked out
  
\ba
C_{(n_{1}n_{2})(n_{3}n_{4})}(t)=\sum_{j m}I_{n_{2}n_{4}}(t)I_{n_{3}n_{1}}(t)/\hat j^{4}
\label{eq14}
\ea  

with  
  
\ba
I_{n_{3}n_{1}}(t)=\langle n_{3}| e^{-it\omega_{s} a^{+}a} e^{-i\eta j(j+1)\sin(\omega_{s}t)a^{+}}                                         
e^{-i\eta j(j+1)\sin(\omega_{s}t)a}
\notag \\
e^{\omega_{s}[1-\cos(\eta j(j+1)t)]a^{+}/j(j+1)\eta}
e^{-\omega_{s}[1-\cos(\eta j(j+1)t)a/j(j+1)\eta]}|n_{1}\rangle e^{\Psi(t)} 
\label{eq15}
\ea   

and
  
\ba
I_{n_{2}n_{4}}(t)=\langle n_{2}| e^{+it\omega_{s} a^{+}a} e^{+i\eta j(j+1)\sin(\omega_{s}t)a^{+}}                                         
e^{+i\eta j(j+1)\sin(\omega_{s}t)a}
\notag \\
e^{\omega_{s}[1-\cos(\eta j(j+1)t)]a^{+}/j(j+1)\eta}
e^{-\omega_{s}[1-\cos(\eta j(j+1)t)]a/j(j+1)\eta }|n_{4}\rangle e^{\Psi(t)}
\label{eq16} 
\ea 
The matrix elements of $\hat \rho_{S}(t)$ read

\ba
\rho^{n_{3}n_{4}}_{S}(t)=\sum_{n_{1},n_{2}}c_{n_{1}}c^{*}_{n_{2}}C_{(n_{1}n_{2})(n_{3}n_{4})}(t)
\label{eq17}
\ea   
     
The expressions of $I_{n_{3}n_{1}}(t)$ and $I_{n_{2}n_{4}}(t)$ are explicitly worked out in Appendix C by means of the Zassenhaus series development~\cite{zas}, see Appendix A. They are exact.
It can be seen that the expressions are oscillating functions of $t$, hence neither the diagonal nor the non-diagonal elements will decrease to and stay at zero when time flows. The system behaves coherently.\\

\section{Model 3: A non-Markovian system}

Consider the case where the Hamiltonian $\hat H$ of the total system reads
  
\ba
\hat H= \hat H_{S}+\hat H_{E}+\hat H_{SE}
\label{eq18}
\ea

with 

\begin{center}
\ba
\hat H_{S}=\omega \hat J_{z} 
\notag \\
\hat H_{E}=\beta b^{+}b
\notag \\
\hat H_{SE}=\eta(b^{+}+b) \hat J^{2}
\label{eq19}
\ea 
\end{center}  
which corresponds to the case where the role of $S$ and $E$ defined in model $2$ are exchanged, $b^{+},b$ are boson operators, $\omega$ is the rotation frequency of the system, $\beta$ the quantum of energy of the oscillator and 
$\eta$ the strength parameter in the coupling interaction between $S$ and $E$.

Now $\hat C_{H} \not= 0$. Since $\hat J_{z}$ and $\hat J^{2}$ commute in the basis of states $[|j m\rangle]$ the 
matrix elements of $\hat H$ in $S$ space read
  
\ba
\langle jm|\hat H|jm \rangle=\omega m+\beta b^{+}b+\eta j(j+1)(b^{+}+b) 
\label{eq20}
\ea 
  
The expression of the density operator $\hat \rho_{S}(t)$ at time $t$ is then obtained by taking the trace over the environment states of the total Hamiltonian $\hat \rho(t)$ leading to 

\ba
\hat \rho_{S}(t)=Tr_{E}\hat \rho(t) 
\label{eq21}
\ea 
whose matrix elements read  
  
\ba
\rho^{j_{1} m_{1}, j_{2} m_{2}}_{S}(t)=\rho^{j_{1} m_{1}, j_{2} m_{2}}_{0}(t)\Omega_{E}(j_{1},j_{2,}t)
\label{eq22}
\ea 
with  
  
\ba
\rho^{j_{1} m_{1}, j_{2} m_{2}}_{0}(t)=\frac{e^{[-i\omega(m_{1}-m_{2})]t}}{(\hat j_{1}\hat j_{2})^{1/2}}
\label{eq23} 
\ea  
with $\hat j_{i}=2j_{i}+1$. The bosonic environment contribution can be put in the following form 
  
\ba
\Omega_{E}(j_{1},j_{2,}t)=\sum_{n=0}^{n_{max}}\frac{1}{n!}\sum_{n',n^{"}}\frac{E_{n,n'}(j_{1},t)
E^{*}_{n^{"},n}(j_{2},t)}{[(n'!)(n''!)]^{1/2}}
\label{eq24} 
\ea  
The results are exact. The Zassenhaus development formulated in Appendix A has been used in order to work out the expressions~\cite{zas}. The expressions of the polynomials $E_{n,n'}(t)$ and $E^{*}_{n'',n}(t)$ are developed in Appendix B.\\

By simple inspection of the expressions in Appendix B it can be seen that the non-diagonal of $\rho^{j_{1} m_{1}, j_{2} m_{2}}_{S}(t)$ may cross zero when $t$ increases but oscillate and never reach and stay at zero whatever the length of the time interval which goes to infinity. Hence no decoherence will be observable in this case. 
  
\section{Model 4: A measure model}
 
As a fourth example we consider now a model which was introduced by Cuccoli in the framework of a study of the correlation between measurement and decoherence~\cite{cuc}.

\subsection{The model}  
 
Following the von Neumann  scheme for quantum measurements we consider a system $S$ in contact with an environment 
$E$ considered as a measuring device. The wave function of the total system $S+E$ is separable at time $t=0$ and 
takes the form
 
\ba
|\Psi(0)\rangle=|i\rangle \otimes |\gamma\rangle
\label{eq25} 
\ea     
where $|i\rangle, |\gamma\rangle$  are sets of states in $S$ and $E$ space respectively. The $[|i\rangle]$'s 
constitute a so called preferred basis of states of the system $S$ which has to stay unchanged in time during the measuring operation

\ba
|i\rangle |\gamma\rangle \rightarrow |i\rangle |\gamma^{i}\rangle
\label{eq26} 
\ea     
where $|\gamma^{i}\rangle$ are states which are affected by the presence of $S$ space. Similarly for any linear combination of states $|i\rangle$

\ba
\sum_{i}c_{i}|i\rangle |\gamma \rangle \rightarrow \sum_{i}c_{i}|i\rangle |\gamma^{i}\rangle
\label{eq27} 
\ea 
\\

Introducing explicitly time in this scheme the total wave function at time $t$ is given by 
 
\ba
|\Psi(t)\rangle=\sum_{i}c_{i}|i\rangle |\gamma^{i}(t)\rangle
\label{eq28} 
\ea    
Under the assumption that the measuring device does not change with time, i.e. the operator corresponding to the observable to be measured commutes with the total Hamiltonian $\hat H=\hat H_{S}+\hat H_{E}+\hat H_{SE}$, the expression of the density operator $\rho_{S}(t)$ reads

\ba
\hat \rho_{S}(t)=\sum_{i}|c_{i}|^{2} |i \rangle \langle i|+\sum_{i\neq i^{'},\gamma} c_{i} c^{*}_{i'}
|i \rangle \langle i^{'}|\langle \gamma^{i'}(t)|\gamma^{i}(t)\rangle
\label{eq29} 
\ea  
with 
 
\ba
|\gamma^{i}(t)\rangle=e^{-it\hat H^{i}_{E}}|\gamma \rangle
\label{eq30} 
\ea 
Here 
 
\ba
\hat H^{i}_{E}= \langle  i|\hat H_{S}+\hat H_{E}+\hat H_{SE}| i \rangle
\label{eq31} 
\ea  
is an operator in $E$ space
 
\ba
\hat H^{i}_{E}=\epsilon_{i}1_{E}+\hat H_{E}+\hat H_{SE}^{i}
\label{eq32} 
\ea 
where the $\epsilon_{i}$'s are the eigenvalues of $\hat H_{S}$. It comes out that the diagonal contribution to $\hat \rho_{S}(t)$ is time independent. Hence the time evolution of the density operator is governed by the second term which is non-diagonal. 

Working out the second term in the expression of $\hat \rho_{S}(t)$ for fixed state $|\gamma\rangle$, $|i \rangle$ and  $|i^{'} \rangle$ leads to matrix elements of the form

\ba
M^{i i'}_{\gamma,\gamma'}=c_{i} c^{*}_{i'}|i \rangle \langle i^{'}| e^{it(\epsilon_{i}-\epsilon_{i'})}
e^{it(E_{\gamma}-E_{\gamma^{'}})} \langle \gamma|e^{it \hat H_{SE}^{i}}|\gamma^{'}\rangle
\langle \gamma^{'}|e^{-it \hat H_{SE}^{i}}|\gamma \rangle
\label{eq33} 
\ea  
where $\epsilon_{i}$ and $E_{\gamma}$ are the eigenvalues of $\hat H_{S}$ and $\hat H_{E}$.\\

The central point of interest concerns the behaviour in time of the non-diagonal terms given by Eq.(29) since they signalize the coherent or decoherent behaviour of the system.

We consider two different cases which correspond to two different types of interaction Hamiltonians 
$\hat H_{SE}^{i}$.
 
\subsection{Case a): $[\hat H_{E},\hat H_{SE}^{i}] = 0$, $i \not= i'$} 

This corresponds to a Markovian behaviour of $S$ since the matrix elements $M$ are diagonal in $E$ space
~\cite{sti,riv,kr}. In this case

\ba
M^{i i'}_{\gamma,\gamma}=c_{i} c^{*}_{i'}|i \rangle \langle i^{'}| e^{it(\epsilon_{i}-\epsilon_{i'})}
\langle \gamma|e^{it \hat H_{SE}^{i}}|\gamma\rangle
\langle \gamma|e^{-it \hat H_{SE}^{i^{'}}}|\gamma \rangle
\label{eq34} 
\ea 
i.e. the environment (measuring device) stays in the same state all along its evolution in time. The contributions of the sum over $\gamma$ consist of real phases which add up as a sum of oscillations. A priori this never leads to the annulation of the non-diagonal terms $M^{i i'}_{\gamma,\gamma}$ for any finite or infinite time interval even though
the sum may reach zero at some interval of times. Hence the system remains coherent.  

\subsection{Case b): $[\hat H_{E},\hat H_{SE}] = \kappa \hat I_{E}$, $i \not= i'$}  

Here $\hat I_{E}$ is the identity operator in $E$ space. Using the Zassenhaus decomposition~\cite{zas} the non-diagonal contribution of the super matrix $M$ reads
\ba
M^{i i'}_{\gamma,\gamma'}=c_{i} c^{*}_{i'}|i \rangle \langle i^{'}| e^{it(\epsilon_{i}-\epsilon_{i'})}
e^{it(E_{\gamma}-E_{\gamma^{'}})} \langle \gamma|e^{it \hat H_{SE}^{i}}|\gamma^{'}\rangle
\langle \gamma^{'}|e^{-it \hat H_{SE}^{i}}|\gamma \rangle e^{\kappa t^{2}/2}
\label{eq35} 
\ea   
For the physical case where $\kappa<0$ the non-diagonal matrix elements of $\hat \rho_{S}(t)$ decrease non monotonically to zero with a time scale $1/\kappa$. 

\section{Model 5: Markovian limit of an open system and decoherence}

As a last example we introduce a model proposed in refs.~\cite{wei,heng,wei1} in order to show how decoherence 
can be generated under certain conditions in the Markovian limit of a non-Markovian process. 

\subsection{The model}

The system $S$ is a many-body system of particles with eigenstate energies $[e_{i}, i=1,N]$ coupled to an environment $E$ of non interacting bosons or fermions through an interaction $\hat H_{SE}$ characterized by a spectral density $J_{\alpha i j}$ where the index ${\alpha}$ designates the particles in the environment. The spectral function reads~\cite{wei,wei1}

\ba
J_{\alpha i j}=2\pi\sum_{k}V_{\alpha k i}V^{*}_{\alpha kj} \delta(\omega-\epsilon_{k})
\label{eq36} 
\ea 
where  $V_{\alpha k i}$ is the interaction between the system states and the environment states $|\alpha \rangle$. 
By means of the coherent state path integral techniques ~\cite{zha} in the framework of the Feynman-Vernon influence functional formalism ~\cite{fey} it is possible to derive the master equation which governs the evolution of 
$d \hat \rho_{S}(t)/dt$. This equation has the structure of a Lindblad equation~\cite{gl} with two terms. The second term is governed by time dependent dissipation coefficients $c_{ij}(t)$.They are obtained from the Green's functions 
$G_{ij}(\tau,t_{0})=\langle [a_{j}^{+}(\tau),a_{i}(t_{0})]\rangle$. 



In matrix form

\ba
{\bf c}(\tau,t_{0})=-1/2[d{\bf G}(\tau,t_{0})/dt {\bf G}^{-1}(\tau,t_{0}) + h.c.]
\label{eq37} 
\ea 

\subsection{Solution in a specific case}

The $N*N$ Green's functions ${\bf G}(\tau,t_{0})$ obeys the following equation

\ba
\frac{d}{d\tau} {\bf G}(\tau,t_{0})+i {\bf e_{s}}{\bf G}(\tau,t_{0})+\int_{t_{0}}^{\tau}d\tau{'} {\bf v}(\tau,\tau^{'}){\bf G}(\tau^{'},t_{0})=0
\label{eq38} 
\ea
where ${\bf e}_{s}$ is the diagonal $N*N$ eigenvalue matrix of the states in $S$ and the $N*N$ propagators ${\bf v}(\tau,\tau^{'})$ are given by 

\ba
{\bf v}(\tau,\tau^{'})=\sum_{\alpha}\int \frac{d\omega}{2\pi}{\bf J}_{\alpha}(\omega)\exp(-i\omega(\tau-\tau^{'}))
\label{eq39} 
\ea

In order to be able to work out the solution of eqs.(38) and (39) in terms of explicit algebraic expressions we 
consider a diagonal spectral function ${\bf J}(\omega)$. The $\omega$ dependence of  ${\bf J}$ leads to dissipation functions $c_{ii}(t)$ which are time dependent functions and the behaviour of $\hat \rho_{S}(t)$ is non-Markovian. 

If however ${\bf J}={\bf J_{0}}=J_{0}{\bf 1}>0$ the propagator ${\bf v}(\tau,\tau^{'})= 
{\bf J_{0}}\delta(\tau-\tau^{'})$
and the solution of eq.(38) reads

 \ba
{\bf G}(\tau,t_{0})=\exp{-i({\bf e_{s}}-i{\bf J_{0}})(\tau-t_{0})}
\label{eq40} 
\ea

Introducing this quantity on the r.h.s. of eq.(37) leads to 

\ba
{\bf c}(t)= {\bf J_{0}}
\label{eq41} 
\ea 
hence to a constant quantity. As a consequence the master equation which governs the evolution of $\hat \rho_{S}(t)$ 
gets a Lindblad expression corresponding to a Markovian process. In the general case the density operator possesses non-diagonal elements which monotonously decay to zero with time.

\section{Summary and conclusions: what can one learn from these examples?}

In the present work we investigated the coherence properties of open quantum systems evolving in a finite Hilbert space $S$ and coupled to an environment evolving in a finite Hilbert space $E$. We considered Markovian and non-Markovian systems and characterized this property by means of the structural properties of the different components of the total Hamiltonian of the system and its environment. The investigation was performed by means of five models which encompass different physical cases.

\begin{itemize}

\item The first two models described Markovian systems. In the first one the environment space $E$ contained a unique state ~\cite{kr}. It came out that the system $S$ was constrained to evolve in time in the state in which it was created at the origin of the process. Hence in the initial chosen diagonal basis of states the density operator of the system did not possess non-diagonal matrix elements and, as a consequence, no decoherence process can be observed.      

The second model generalized the former case. It was chosen such that the Hamiltonian of the environment $\hat H_{E}$ 
and the interaction $\hat H_{SE}$ between the system and the environment commute, $[\hat H_{E},\hat H_{SE}]=0$, which characterizes a Markovian process~\cite{riv,kr}. This specific property which corresponds to a less stringent constraint than the one imposed in the first case led to a density operator which possessed also non-diagonal elements. However the non-diagonal elements came out to be oscillating functions of time which again did not lead to a decay to zero with time. 

Such a process is different from the expected behaviour generically obtained from the solution of a Lindblad equation.  

\item The third system described a non- Markovian process which was here defined by the commutation relation  
$[\hat H_{E},\hat H_{SE}] \neq 0$. It came out that there is no decoherence to be observed, the matrix elements of the density operator oscillate in time.

\item The fourth model relied on the von Neumann approach a measure process~\cite{cuc}. There the density operator 
was decomposed into a diagonal and a non-diagonal part with respect to a preferrential basis of states in the system space $S$.
Two cases were examined:

- In the first case where $[\hat H_{E},\hat H_{SE}]=0$ the non-diagonal contributions added up as a sum over the environment states. They may behave incoherently in time and decrease to or close to zero but this decrease is not monotonous,the contribution may again increase over later intervals of time and repeat to infinity.

- In the second case the non-diagonal matrix elements were governed by $[\hat H_{E},\hat H_{SE}]\simeq \hat I$ where $\hat I$ is the unity operator in the total space $S+E$. There one observes an exponential decrease in $t^{2}$ modulated by oscillating contributions. 

\item The fifth model was chosen as a description of the transition from a Markovian to a non-Markovian process. We chose a model already developed in ref.~\cite{wei} which describes the time evolution of an open system coupled to its environment through an interaction given in terms of a spectral function. The master equation which governs the evolution of the density operator was put in the form of a Lindblad equation~\cite{gl,bre,dio}. In the general case the dissipation functions which enter the second term of the equation are time dependent~\cite{hal}, the system follows a non-Markovian process and does not necessarily decohere. However by extending the spectral function to a continuum 
with an infinite number of states the equation went over to a genuine Lindblad equation for the density operator.  

\end{itemize}

From these results it is possible to draw the following conclusions:

\begin{itemize}

\item Markovian processes characterized by the divisibility property leading to $[\hat H_{E},\hat H_{SE}]=0$ 
~\cite{riv,sti,kr} do not necessarily generate a decoherent behaviour of the open system. This result is in agreement with formal previous work, see~\cite{agr,lid1,lid2}.

\item A non-Markovian evolution which is characterized by  $[\hat H_{E},\hat H_{SE}]\neq 0$ may or may not lead to a coherent or decoherent process. The nature of the evolution depends on the structure of the evolution operator
$e^{it\hat H}$ where $\hat H$ is the total Hamiltonian of $S,E$ and their interaction, i.e. the commutation properties of the Hamiltonians of the system, the environment and the interaction which acts between them.

\item The transition from a Markovian to a non-Markovian process can be induced by the nature of the energy spectrum of the environment $E$. In the model which was worked out ~\cite{wei,wei1,kr} the interaction was governed by a spectral function which has an infinite energy extension. In this limit the evolution gets Markovian, it is governed by a Lindblad equation which leads to a monotonous decoherence process whereas a non-Markovian process generates different regimes which oscillate or not to zero ~\cite{heng}.  

\end{itemize}

\section{Appendix A: the Zassenhaus development} 
 
If $X=-i(t-t_{0})(\hat H_{S}+\hat H_{E})$ and $Y=-i(t-t_{0})\hat H_{SE}$

\ba
e^{X+Y}=e^{X}\otimes e^{Y}\otimes e^{-c_{2}(X,Y)/2!}\otimes e^{-c_{3}(X,Y)/3!}\otimes e^{-c_{4}(X,Y)/4!}...
\label{eq31}
\ea

where

\begin{center}
$c_{2}(X,Y)=[X,Y]$\\ 
$c_{3}(X,Y)=2[[X,Y],Y]+[[X,Y],X]$\\ 
$c_{4}(X,Y)=c_{3}(X,Y)+3[[[X,Y],Y],Y]+[[[X,Y],X],Y]+[[X,Y],[X,Y]$, etc.\\
\end{center} 
 
The series has an infinite number of term which can be generated iteratively in a straightforward way ~\cite{ca}. If $[X,Y]=0$ the truncation at the  third term  leads to the factorisation of the $X$ and the $Y$ contribution. If $[X,Y]=c$ where $c$ is a c-number the expression corresponds to the well-known Baker-Campbell-Hausdorff formula.\\  

Remark: the Zassenhaus expansion has a finite range of convergence. The upper convergence limit of time is in principle given by~\cite{suz} 
  
\ba
t=1/2\ln 2/(\|\hat H_{E}\|+\|\hat H_{SE}\|)
\ea  
Here the different models which are developed are analytically integrable, hence the series can be formally summed up to infinity.

\section{Appendix B: The bosonic content of the density operator}
 
The expressions of the bosonic contributions to the density matrix $\rho^{j_{1} m_{1}, j_{2} m_{2}}_{s}(t)$ are given by 
 
\ba
E_{n,n'}(j_{1},t)=e^{-i\beta t}\sum_{n\geq n_{2},n_{3}\geq n_{2}}\sum_{n_{3}\geq n_{4},n'\geq n_{4}}(-i)^{n+n_{3}}
(-1)^{n'+n_{2}-n_{4}}
\notag\\
\frac{n!n'!(n_{3}!)^{2}[\alpha(t)^{n+n_{3}-2n_{2}}][\zeta(t)^{n_{3}+n'-2n_{4}}]}{(n-n_{2})!(n_{3}-n_{4})!
(n_{3}-n_{2})!(n'-n_{4})!}e^{\Psi_{1}(t)}
\label{eq23}      
\ea
and

\ba
E^{*}_{n^{"},n}(t;j_{2})=e^{i\beta t}\sum_{n^{"}\geq n_{2},n_{3}\geq n_{2}}\sum_{n_{3}\geq n_{4},n\geq n_{4}}i^{n^{"}+n_{3}}
(-1)^{n+n_{2}-n_{4}}
\notag\\
\frac{n^{"}!n!(n_{3}!)^{2}[\alpha(t)^{n^{"}+n_{3}-2n_{2}}][\zeta(t)^{n+n_{3}-2n_{4}}]}{(n^{"}-n_{2})!(n_{3}-n_{2})!(n_{3}-n_{4})!(n-n_{4})!}e^{\Psi_{2}(t)}
\label{eq24}      
\ea

The different quantities which enter $E_{n,n'}(t)$ are 

\ba
\alpha(t)=\frac{\gamma(j_{1})\sin\beta t}{\beta}
\label{eq25}      
\ea

\ba
\zeta(t)=\frac{\beta[1-\cos\gamma(j_{1})t]}{\gamma(j_{1})}
\label{eq18}      
\ea

\ba
\gamma(j_{1})=\eta j_{1}(j_{1}+1)
\label{eq26}      
\ea

\ba
\Psi_{1}(t)=-\frac{1}{2}[\frac{\gamma^{2}(j_{1})\sin^{2}(\beta t)}{\beta^{2}}+\frac{\beta^{2}(1-\cos\gamma(j_{1})t)^{2}}{\gamma^{2}(j_{1})}]                           
\label{eq27}      
\ea

and for $E^{*}_{n'',n}(t)$: 
\ba
\alpha(t)=\frac{\gamma(j_{2})\sin\beta t}{\beta}
\label{eq28}      
\ea

\ba
\zeta(t)=\frac{\beta[1-\cos\gamma(j_{2})t]}{\gamma(j_{2})}
\label{eq29}      
\ea

\ba
\gamma(j_{2})=\eta j_{2}(j_{2}+1)
\label{eq30}    
\ea

\ba
\Psi_{2}(t)=-\frac{1}{2}[\frac{\gamma^{2}(j_{2})\sin^{2}(\beta t)}{\beta^{2}}+\frac{\beta^{2}(1-\cos\gamma(j_{2})t)^{2}}{\gamma^{2}(j_{2})}]                           
\label{eq31}      
\ea

\section{Appendix C}
 
The matrix elements $I_{n_{3}n_{1}}(t)$ and $I_{n_{2}n_{4}}(t)$  read 
 
\ba
I_{n_{3}n_{1}}(t)=e^{-i\omega_{s}t}\sum_{m_{2}m_{3}m_{4}}(-1)^{n_{1}+m_{2}+m_{4}}(-i)^{n_{3}+m_{3}}
d_{1}(t)^{n_3+m_{3}-2m_{2}}
\notag \\
d_{2}(t)^{n_1+m_{3}-2m_{4}}\frac{n_{3}!n_{1}!(m_{3}!)^{2}e^{\Psi(t)}}{(n_{3}-m_{2})!(m_{3}-m_{2})!(m_{3}-m_{4})!
(n_{1}-m_{4})!}
\label{eq32}
\ea   

and
  
\ba
I_{n_{2}n_{4}}(t)=e^{+i\omega_{s}t}\sum_{p_{2}p_{3}p_{4}}(-1)^{p_{2}+p_{4}+n_{4}}(+i)^{n_{2}+p_{3}}
d_{1}(t)^{n_2+p_{3}-2p_{2}}
\notag \\
d_{2}(t)^{n_4+p_{3}-p_{4}}\frac{n_{2}!n_{4}!(p_{3}!)^{2}e^{\Psi(t)}}{(n_{4}-p_{4})!(p_{3}-p_{2})!(p_{3}-p_{4})!
(n_{2}-p_{2})!}
\label{eq33} 
\ea      

For fixed $n_{1},n_{2},n_{3},n_{4}$ the sums over $({m_{i},p_{i}})$ indices run over all values such that all factorials are positive numbers or zero.

The coefficients $d_{1}(t), d_{2}(t)$  read

\ba
d_{1}(t)=\eta j(j+1)\sin(\omega_{s}t)/\omega_{s}
\label{eq34}
\ea   

\ba
d_{2}(t)=\omega_{s}[1-\cos(\eta j(j+1)t)]]/\eta j(j+1)
\label{eq35}
\ea   
and  
 
\ba
\Psi(t)=-\eta^{2}[j(j+1)]^{2}\sin^{2}(\omega_{s}t)/2(\omega_{s}^{2})-\omega_{s}^{2}[1-\cos(\eta j(j+1)t)]^{2}
/2(j(j+1))^{2}\eta^{2}
\label{eq36}
\ea

\end{document}